# Optical attenuators extend dynamic range but alter angular response of planar ultraviolet-C dosimeters

Alison Su[1#], Alisha Geldert[1#], Samantha M. Grist[2], & Amy E. Herr[1,2*]
[#]These authors contributed equally.
*Corresponding author: aeh@berkeley.edu
[1]University of California, Berkeley-University of California, San Francisco Graduate Program in Bioengineering, [2]Department of Bioengineering, Berkeley, California 94720, United States

**Abstract:**

A challenge for sensors used in ultraviolet-C (UV-C) decontamination protocols of N95 respirators is validation that the entire N95 surface receives the minimum acceptable dose. Photochromic indicators (PCIs) can accurately measure UV-C dose on nonplanar surfaces, but often saturate below doses required to decontaminate porous, multilayered textiles such as N95s. Here, we investigate the use of optical attenuators to extend PCI dynamic range while maintaining a near-ideal angular response – critical for accurate measurements when UV-C is uncollimated. Through an analytical model, we show that tuning attenuator refractive index, attenuation coefficient, and thickness can extend dynamic range, but compromises ideal angular response unless the attenuator is an ideal diffuser. To demonstrate this tradeoff empirically, we pair PCIs with model specular (floated borosilicate) and diffuse (polytetrafluoroethylene) attenuators, characterize the angular response, and evaluate on-N95 UV-C dose measurement accuracy of each PCI-attenuator stack in a UV-C decontamination system. While both borosilicate and polytetrafluoroethylene increase PCI dynamic range >4×, both attenuators introduce angle-dependent transmittance, which causes location-dependent underestimation of UV-C dose. The PCI-borosilicate and PCI-polytetrafluoroethylene stacks underreport true on-N95 dose by 1) 14.7% and 3.6%, respectively, on a surface near-normal to the array of source lamps, and 2) 40.8% and 19.8%, respectively, on a steeply sloped location. Overall, we demonstrate that while planar optical attenuators can increase PCI dynamic range, verification of near-ideal angular response is critical for accurate UV-C dose measurement.

**Introduction**

Ultraviolet-C (UV-C) radiation is a key germicidal technique regularly applied in healthcare settings to decontaminate air[1,2], surfaces[3,4], and recently, N95 respirators to address the COVID-19 pandemic-induced shortages[5,6]. UV-C photons catalyze protein and nucleic acid photodegradation; after sufficient cumulative photon absorption (UV-C dose), compromised pathogens are inactivated. The UV-C dose needed for decontamination depends on the pathogen, substrate, and other factors[7]. In particular, porous and multilayered textiles such as N95 respirators and surgical masks and gowns require higher applied outer surface doses as compared to nonporous materials, to offset attenuation of UV-C reaching pathogens embedded in the inner material layers[8,9]. Decontamination efficacy is directly related to UV-C dose, and UV-C dose measurements are frequently the only metric bridging laboratory viral inactivation studies and clinical implementation; thus, accurate UV-C dose measurements are critical for protocol validation.

Validation of decontamination of N95s and other porous and/or nonplanar substrates poses unique UV-C measurement challenges. The ~100× higher UV-C dose required to decontaminate porous materials as compared to nonporous surfaces[3,8,10] require UV-C sensors with sufficiently high dynamic range. UV-C systems often deliver nonlinear doses over time[11,12], precluding extrapolation from short exposures. Additionally, the complex N95 geometry complicates measurement accuracy, as the UV-C dose received by a surface at a given angle of incidence $\theta$ is reduced by a factor of $\cos(\theta)$ from the dose received at normal incidence (Lambert's cosine law[13]). Thus, UV-C dose measurement accuracy depends on how proportional the sensor readout over angles of incidence $0° \leq \theta \leq 90°$ (termed "angular response") is to $\cos(\theta)$ (termed "ideal response"). A sensor with an ideal response is critical for applications such as N95 decontamination, which involves both nonplanar targets and uncollimated UV-C. However, sensor housing, spectral filters, and other elements in the optical path often alter angular response[14] and sensor angular response is often non-ideal[15,16], uncharacterized, or unreported.

UV-C photochromic indicators (PCIs), which change color in response to UV-C dose, overcome many challenges associated with on-N95 measurements. PCIs can have an ideal angular response[17] because PCI dose response and specificity are governed by chemistry[18] rather than additional physical elements within the optical path. Though PCI readout is traditionally qualitative or at-best semi-quantitative (if a color swatch to dose reference is provided), a recent study developed a robust workflow to quantify UV-C dose from PCI color change to map UV-C dose across N95 facepieces[11]. However, because PCIs were originally designed to validate nonporous surface decontamination, UV-C doses required for porous material decontamination typically exceed the PCI dynamic range. Thus, an extended PCI dynamic range spanning higher UV-C doses is urgently needed to validate decontamination of porous materials like N95s.

There are two approaches to extend the PCI dynamic range: (1) altering the chemistry governing the PCI color change, (e.g., adding reagents to modify the reaction kinetics or equilibrium[18,19]), or (2) attenuating UV-C incident on the PCI[20,21]. As a PCI-agnostic approach, attenuation lends itself to widespread adoption across diverse settings. However, objects within the optical path may alter the PCI angular response due to angle-dependent refraction, reflection, scattering, and absorption[14,22]. A non-ideal angular response will cause angle-dependent dose measurement errors. If the angle of incidence is known or constant, an angle-dependent correction factor can be determined[11,23,24]. However, the deformable N95 facepiece shape combined with significant UV-C scattering and reflection render this correction-factor approach infeasible for N95 UV-C decontamination systems.

Here, we employ theoretical and empirical approaches to investigate whether readily available materials can serve as optical attenuators to extend PCI dynamic range while maintaining measurement accuracy for N95 decontamination protocol validation. We develop an analytical model based on fundamental optics principles and attenuator properties to predict attenuator transmittance as a function of angle of incidence. Analytically and empirically with a point-like UV-C source, we characterize the angular response of PCIs stacked directly behind (with respect to the optical axis) each of two model attenuator materials: one non-diffuse and one diffuse. Finally, to mimic implementation in an N95 decontamination protocol, we evaluate the measurement accuracy of each PCI-attenuator stack on two differently sloped N95 facepiece locations in a decontamination chamber, where UV-C angles of incidence are unknown. We demonstrate that although attenuators with diffuse properties improve angular response as compared to non-diffuse attenuators, a model planar diffuse attenuator still alters angular response, which compromises measurement accuracy. In total, we develop frameworks to relate key material properties of optical attenuators to the dynamic range and angular response of the PCI-attenuator stack and assess model PCI-attenuator stacks in an example end-use case to highlight critical considerations when modifying planar dosimeters for measurements on nonplanar surfaces.

**Materials and Methods**
Materials

The attenuators used were floated borosilicate (Borofloat®, 25.4 mm width × 25.4 mm length × 1.1 mm ± 0.1 mm thickness, 80/50 scratch/dig quality, Precision Glass & Optics 0025-0025-0011-GE-CA), referred to as "borosilicate", and polytetrafluoroethylene film (Teflon®, 0.51 mm thickness, cut into 25.4 mm squares, McMaster-Carr 8569K23), referred to as "PTFE". All radiometer measurements were collected using a calibrated ILT1254 UV-C radiometer with a Teflon dome diffuser (International Light Technologies). PCIs were UVC 100 Dosimeter dots (American Ultraviolet). For transmittance and angular response measurements, a modified handheld UV-C lamp (EF-140) with one BLE-2537S amalgam bulb (254 nm emission) and a UV-C-blocking plate with a 25.4 mm-diameter aperture installed was used as a point-like UV-C source (Spectronics). PCI and PCI-attenuator stack calibration curves and on-N95 measurements were made in a commercial UV-C decontamination chamber (Spectronics XL-1000 UV-C with an array of 5 BLE-8T254 254 nm low-pressure amalgam bulbs along the top) with a small custom notch in the door for the radiometer cord to pass through. All on-N95 measurements were made on one 3M 1860 N95 respirator.

All analytical modelling and analysis were performed in MATLAB® R2020b.

Borosilicate transmittance measurement

To measure total transmittance through borosilicate, a radiometer placed normal to the point-like UV-C source at a distance of 127 mm recorded the irradiance with and without borosilicate in the optical path (Figure S1A).

Analytical model

The attenuation coefficient (α) of borosilicate was calculated from the total transmittance measured at near-normal angles of incidence (Figure S1A). We estimated the refractive index $n_{att}$ ≈ 1.50 at 254 nm for borosilicate based on linear extrapolation of $n$ for the two shortest wavelengths reported[25] (~365 nm and 405 nm). We estimated $n_{att}$ ≈ 1.38 for PTFE, as reported by a manufacturer[26].

PCI quantification

PCIs were quantified as previously described[11]. Briefly, D65/10° L*a*b* values of PCIs were measured using an RM200QC spectrocolorimeter (X-rite®). Color change with respect to an unexposed PCI was quantified using the CIEDE2000 ΔE formula[11,27]. To generate calibration curves, a radiometer and PCI were positioned within the UV-C chamber at planar locations of equal irradiance (Figure S2) to measure UV-C dose and CIEDE2000 ΔE, respectively. CIEDE2000 ΔE values and corresponding UV-C doses were fit to a function based on first-order reaction kinetics[11]. Unless otherwise noted, reported errors are the root-sum-square of standard deviations corresponding to both replicate variation and PCI quantification uncertainty.

Angular response measurements with apertured UV-C source

The angular responses of PCI-attenuator stacks were determined from the dose measured by PCIs rotated around an optical post to expose the PCIs to different angles of incident UV-C (0°-90° in 15° increments) from a point-like source, in accordance with the range of angles used to characterize other dosimeters[28] (Figure S1B). To ensure the UV-C source was point-like, we placed the PCI-attenuator stack in line with the UV-C source, at a distance where source power was independent of distance. UV-C source power (calculated from the Keitz formula from radiometer-measured irradiance) was determined to be independent of distance (i.e., varied by <5% between distances) at distances ≥102 mm using the method previously described[12,17].

To ensure consistent UV-C source output across measurements, dose was monitored using a radiometer at an offset, non-shadowed location; all PCIs within an angular response set were exposed to the same radiometer-measured dose. After UV-C exposure, the PCI-attenuator stack was disassembled and dose received by the PCI was immediately determined ("PCI quantification").

On-N95 dose measurements with PCI-attenuator pairs

On-N95 dose measurements were made at two N95 facepiece locations: near the apex where the N95 surface is nearly normal to the UV-C bulb array ("low-angle"), and near the base where the N95 surface is steeply sloped ("high-angle"). For consistent placement, high- and low-angle locations were marked on the N95, and facepiece deformation was minimized. During each exposure, the N95 was centered in the UV-C chamber, and a radiometer at a fixed location in the chamber recorded irradiance. A chamber floor map reduced positioning error (Figure S3).

To measure on-N95 dose with PCI-attenuator stacks, PCIs were taped with the sensor side flush against the attenuator. The PCI-attenuator stack was then attached to the N95 facepiece using double-sided tape. Measured on-N95 dose was determined from PCI-attenuator calibration curves generated within the UV-C chamber. To compare to the bare PCI results, PCI-attenuator calibration curves were generated from the same locations in-chamber. To evaluate the accuracy of on-N95 PCI-attenuator stack measurements, we calculated true on-N95 dose by multiplying the *in-situ* radiometer measurement by the ratio of irradiance at each on-N95 location to the radiometer location. Irradiance ratios were predetermined using bare PCIs exposed to lower doses, within the bare PCI dynamic range (Note S1)

**Results & Discussion**
**Design specifications relevant to pathogen inactivation**

In this study, we sought to characterize the performance of PCIs stacked behind optical attenuators in measuring UV-C surface doses required for viral inactivation throughout porous materials on nonplanar N95 facepieces. Because planar materials are accessible and scalable (can be cut to size from bulk material), we chose to study planar attenuators. We identified key performance specifications relevant to measurement accuracy: dynamic range and angular response (Figure 1A). We define the PCI dynamic range[11] as the UV-C doses between a lower and upper limit of quantification (LLOQ and ULOQ, respectively) where the relative PCI quantification uncertainty is <10% (Figure S4). As studies support ≥1.0 J/cm$^2$ for ≥99.9% inactivation of non-enveloped viruses on most N95 models[29–31], the PCI-attenuator stack ULOQ must exceed 1.0 J/cm$^2$ for N95 decontamination protocol validation. However, pathogen- and model-specific UV-C efficacy may require higher ULOQ, and should be determined on a case-by-case basis. Additionally, on-N95 dose has been found to vary by ~20× within a decontamination system[11]. To maximize the continuous measurement range in order to characterize the full range of nonuniform doses within a system, the PCI-attenuator stack LLOQ must remain below the bare PCI ULOQ (0.261 J/cm$^2$ for the PCI model and color-readout method used here[11]; Figure S4).

UV-C dose measurement accuracy on nonplanar surfaces depends on the angular response of the detector. Depending on attenuator material properties, transmittance may change with angle of incidence due to angle-dependent reflection, absorption, and degree of scattering (i.e., specular or diffuse reflectance and transmittance), leading to a non-ideal angular response. Because non-ideal angular response is infeasible to correct for without prior knowledge of the angle(s) of incidence, we sought to identify a PCI-attenuator stack with near-ideal angular response. At a given angle of incidence θ, deviation from the ideal angular response is defined as the cosine error[32] (Eq. 1). Integration of the cosine error between 0° and 80° (integrated cosine error, Eq. 2, defined[32] in ISO/CIE 19476) quantifies the overall deviation from the ideal angular response[33]:

$$\text{Cosine error} = f_2(\theta) = \left(\frac{response(\theta)}{response(0°) \cdot cos(\theta)} - 1\right) \times 100\% \tag{1}$$

$$\text{Integrated cosine error} = \int_0^{80°} |f_2(\theta)| \cdot sin(2\theta) d\theta \tag{2}$$

To match the order of magnitude of bare PCI measurement error[11,12] (average error of 7%), PCI-attenuator stack cosine error magnitude must remain ≤10% over all angles of incidence (0-90°).

**Optical properties governing attenuator design for measurements on non-planar surfaces**

To inform design of an attenuator that meets the required specifications, we first sought to identify and relate optical properties that affect attenuator transmittance through a planar material. Transmittance will affect both the dynamic range and angular response of a PCI-attenuator stack. Attenuators may exhibit entirely specular reflection and transmission (i.e., no scattering effects, 'non-diffuse'), or diffuse scattering at the interface ('surface diffusers'), within the material ('volume diffusers'), or at both the interface and throughout the material. We developed an analytical model for total transmittance ($T_{total}$) through materials based on two main interactions (Eq. 3): (1) reflection and refraction at air-attenuator interfaces, which govern the transmittance across the interfaces ($T_{int1}$ and $T_{int2}$) and (2) attenuation throughout the attenuator thickness, which governs the transmittance through the attenuator volume ($T_{mat}$).

$$T_{total} = T_{int1} \cdot T_{mat} \cdot T_{int2} \tag{3}$$

At each air-attenuator interface, the Fresnel equations[14] (Eq. 4) for randomly polarized radiation describe $T_{int}$ based on the air and attenuator refractive indices ($n_{air}$ and $n_{att}$, respectively) and angle of incidence with respect to the surface normal ($\theta_{air}$). Snell's law[34] (Eq. 5) governs the angle of refraction within the attenuator ($\theta_{att}$) (Figure 1B).

$$T_{int} = 1 - \left\{\frac{1}{2} \cdot \left[\left(\frac{n_{air}cos(\theta_{air}) - n_{att}cos(\theta_{att})}{n_{air}cos(\theta_{air}) + n_{att}cos(\theta_{att})}\right)^2 + \left(\frac{n_{air}cos(\theta_{att}) - n_{att}cos(\theta_{air})}{n_{air}cos(\theta_{att}) + n_{att}cos(\theta_{air})}\right)^2\right]\right\} \tag{4}$$

$$n_{air}sin(\theta_{air}) = n_{att}sin(\theta_{att}) \tag{5}$$

Note that the attenuator-to-air interface transmittance ($T_{int2}$) calculation requires interchanging $n_{air}$ and $n_{att}$ as well as $\theta_{att}$ and $\theta_{air}$ in Eq. 4. Specular reflectors have a microscopically flat interface, such that a collimated UV-C beam will strike the material at a single $\theta_{air}$ that governs $T_{int}$. In contrast, due to interface roughness on surface diffusers, the surface normal varies randomly over distances much smaller than the length scale of interest (e.g., dimensions of the PCI)[14]. Thus, the textured interface causes collimated UV-C at any angle to actually strike the microscopically textured interface over a range of $\theta_{air}$. As a result, the proportion of UV-C transmitted across a surface diffuser interface does not depend on the angle of incidence (Figure 1B).

Using this analytical framework, we modeled specular and diffuse interface transmittance as a function of both refractive index difference ($\Delta n$, Figure 1C) and the angle of incidence ($\theta_{air}$, Figure 1D). Increasing $\Delta n$ decreases $T_{int1}$, thus extending the dynamic range of the PCI-attenuator stack (Eq. 4; Figure 1C). To characterize the effect of $\Delta n$ on angular response, we evaluated $T_{int1}$ normalized to $T_{int1}(0°)$ as a function of $\theta_{air}$ over varying $\Delta n$ values (Figure 1D). Because $n$ of most materials[35] is $\leq 2$ and $n_{air} \approx 1$, we evaluated $\Delta n \leq 1$. Surface diffusers exhibit angle-independent transmittance at the interface regardless of $\Delta n$. However, interfaces with specular reflection and transmission have increasingly angle-dependent transmittance as both $\theta_{air}$ and $\Delta n$ increase within the range of values modeled.

Internal transmittance through the attenuator thickness (d) depends on two parameters: the material attenuation coefficient ($\alpha$) and the optical path length through the material (L). Bouguer's law[34] relates $T_{mat}$ to $\alpha$ and L (Eq. 6):

$$T_{mat} = e^{(-\alpha L)} \tag{6}$$

In non-diffuse materials and surface diffusers with no internal scattering, L is dependent on d and $\theta_{att}$ (Eq. 7):

$$L = \frac{d}{\cos(\theta_{att})} \tag{7}$$

In volume diffusers, microstructures within the material scatter rays in random directions[36], decoupling L from $\theta_{att}$. Thus, in volume diffusers, $T_{mat}$ is independent of angle of incidence (Figure 1B).

To elucidate contributions of attenuator properties ($\alpha$ and d) to the magnitude and angle-dependence of $T_{mat}$, we modeled $T_{mat}$ as a function of a nondimensional parameter $\alpha d$ (Figure 1E) and $\theta_{att}$ (Figure 1F). Increasing d or $\alpha$ decreases transmittance via increased material attenuation, thereby extending the PCI dynamic range (Figure 1E). For UV-C transmittance through volume diffusers at any angle, $T_{mat}/T_{mat}(0°)$ is independent of angle of incidence regardless of $\alpha d$. However, increasing $\alpha d$ for non-diffuse materials increases angular dependence of transmittance because 1) increasing d expands the range of optical path lengths over which attenuation occurs, and 2) increasing $\alpha$ increases the sensitivity of $T_{mat}$ on varying path lengths (Figure 1F).

Since the irradiance incident on the PCI-attenuator stack follows Lambert's cosine law[13], the irradiance ultimately incident on the PCI is proportional to $T_{total} \cdot cos(\theta_{air})$. Thus, PCIs stacked directly behind planar attenuators (relative to the optical path) will maintain an ideal response only if $T_{total}$ remains constant over 0° ≤ $\theta_{air}$ < 90°. However, the parameters (Δ$n$, d, and α) required to reduce attenuator transmittance and thus increase the dynamic range of the PCI-attenuator stack concomitantly introduce angle-dependent transmittance. Thus, unless the attenuator diffuses UV-C sufficiently to transmit UV-C independent of angle, there is a fundamental tradeoff between reducing transmittance to extend dynamic range and maintaining an ideal cosine angular response.

**Model diffuse and non-diffuse materials extend the PCI dynamic range beyond 1.0 J/cm²**

To investigate how attenuator material properties affect UV-C dose quantification accuracy, we chose to characterize the performance of PCIs stacked behind each of two widely accessible materials with different degrees of diffuse scattering. Floated borosilicate ("borosilicate") has been demonstrated[11] to extend PCI dynamic range on planar surfaces by ~5×, and thus was chosen as a model non-diffuse attenuator (i.e., exhibits specular reflection and transmission). Polytetrafluoroethylene ("PTFE") was chosen as a model volume diffuser[37], as PTFE is commonly used to improve angular response of radiometers within the ultraviolet range[38,39]. We generated calibration curves for PCIs and PCI-attenuator stacks to verify that chosen attenuator thicknesses extend the PCI dynamic range beyond 1.0 J/cm² (Figure 1G, Figure S4). The bare PCI ULOQ was 0.261 J/cm², below the 1.0 J/cm² design specification for on-N95 dose validation and in line with previous studies[11]. We found that 0.51 mm-thick PTFE and 1.1 mm-thick borosilicate increased the ULOQ to 1.259 J/cm² and 1.853 J/cm², respectively, thus meeting the dynamic range specification. While we only studied one batch of each attenuator, transmittance may vary by batch and should be characterized prior to implementation.

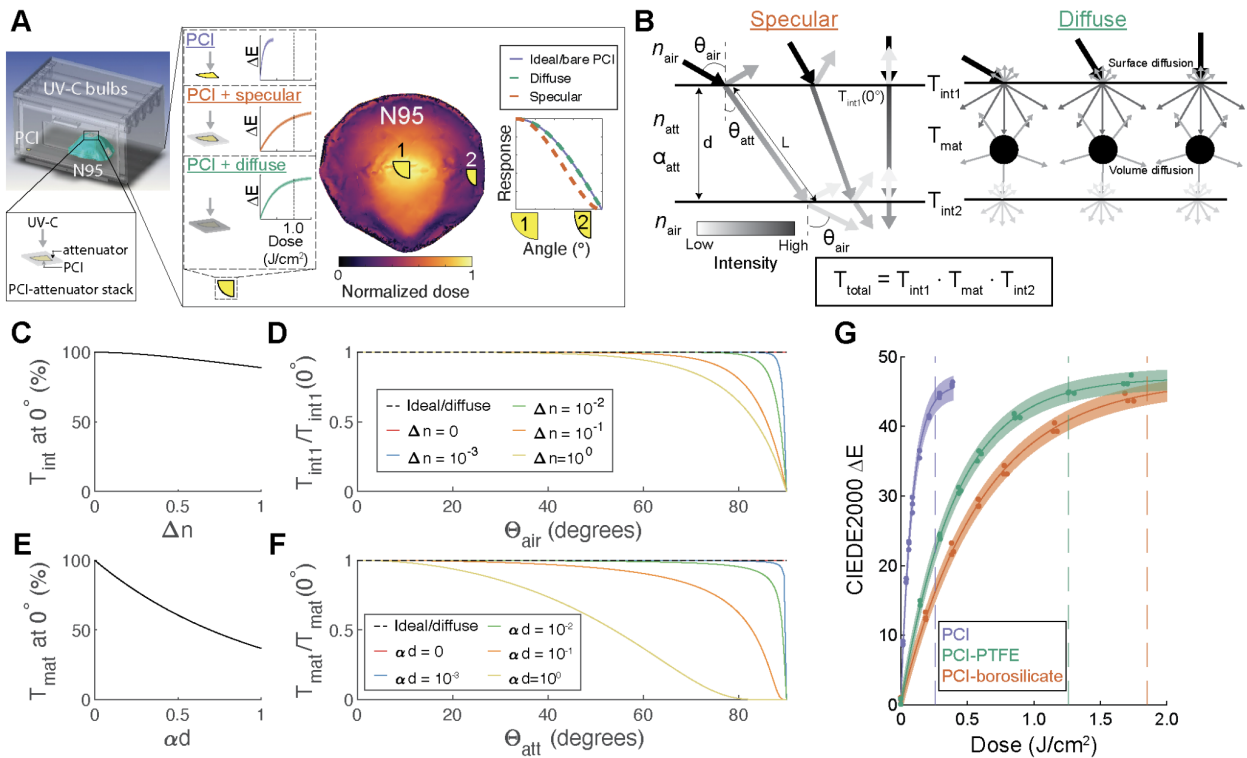

**Figure 1. Attenuator material properties govern dynamic range and angular response of PCI-attenuator stacks.** (A) 3D rendering of N95 UV-C decontamination system with 2D top-down view of chamber floor. Attenuators stacked

in front of PCIs can extend the PCI dynamic range to measure on-N95 dose variation (shown as heatmap), but measurement accuracy on non-planar surfaces like N95s requires an ideal PCI-attenuator angular response. (B) Schematic representation of UV-C transmittance through ideal specular and diffuse attenuators at varying angles of incidence: UV-C enters through the air-attenuator interface ($T_{int1}$), traverses the attenuator ($T_{mat}$), and exits via the attenuator-air interface ($T_{int2}$). Arrow shade represents irradiance magnitude. In non-diffuse materials, reflection and attenuation increase with angle of incidence. In ideal diffusely transmitting materials, transmittance is independent of angle of incidence due to surface and/or volume diffuser behavior. (C-D) Non-zero $\Delta n$ yields both decreased (C) and angle-dependent transmittance (D) at a specular interface. (E-F) Material thickness and attenuation coefficient yield both decreased (E) and angle-dependent (F) transmittance in a non-diffuse material. (G) Two attenuator materials, borosilicate (specular) and PTFE (diffuse) extend the PCI upper limit of quantification beyond 1.0 J/cm$^2$.

## Analytical and empirical characterization demonstrate non-ideal angular response of a model non-diffuse attenuator

To assess quantification accuracy of the PCI-borosilicate stack at different angles of incidence, we compared both the analytical and empirical angular response of a PCI stacked behind 1.1 mm-thick borosilicate to an ideal response. Using an apertured UV-C lamp to achieve near-normal angles of incidence (Figure S5), we measured a total transmittance ($T_{total}$) of 15.63% ± 0.06% for 1.1 mm-thick borosilicate (standard deviation of 3 replicates). We used this thickness and measured $T_{total}$ to predict the angular response of the PCI-borosilicate stack analytically, and also measured angular response with the point-like UV-C source.

As a non-diffuse material, we hypothesized that the PCI-borosilicate stack would readout lower UV-C doses than expected from Lambert's cosine law, with deviations from ideal increasing with angle of incidence due to angle-dependent reflection and absorption[22] (Figure 2A). We calculated the integrated cosine error (Eq. 2) using an upper limit of integration of 75º, the last angle measured <80º. For the PCI-borosilicate stack, we predicted analytically and measured an integrated cosine error of 12.7% and 14.5%, respectfully. Both analytically and empirically, we observed that the UV-C dose transmitted through borosilicate to the PCI underestimates an ideal angular response (Figure 2B). To quantify the deviation from the ideal response as a function of angle, we calculated the cosine error (Eq. 1, Figure 2C). At angles of incidence of 15º and 75º, our model predicted cosine errors of -2.7% and -64.8%, respectively, and we measured cosine errors of -8.2% ± 3.0% and -82.9% ± 5.7%, respectively. Thus, the PCI-borosilicate stack deviated more from an ideal response at higher angles of incidence (Figure 2C), as hypothesized. Importantly, PCI-borosilicate only meets the angular response design specification (i.e., magnitude of cosine error ≤10%) at near-normal angles of incidence: 0º (due to normalization) and 15º empirically, and up to ~29º analytically. While angle-specific correction factors have been determined and applied in tightly controlled systems[24], this approach is not feasible when the distribution of angles of incidence is not precisely known. For N95s in a UV-C chamber, both the 3D N95 facepiece morphology and uncollimated radiation confound application of an angle-specific correction factor to adjust inaccurate on-N95 UV-C dose measurements.

To evaluate the agreement between the analytical model and experiment, we compared the empirical angular response to model predictions. At 3 out of 6 non-normal angles measured, empirical angular response was within error (total propagated error of PCI quantification uncertainty and replicate variation) of model predictions (Figure S6A-B). The difference between empirical and analytical angular response was most substantial at 15° and 75° (Figure S6B), where the empirical normalized angular response was 0.0531 ± 0.0291 and 0.0469 ± 0.0147 below the model predictions, respectively. We hypothesize that the discrepancy between the empirical and analytical angular response arises from error in model parameters (e.g., refractive index, $T_{total}$ at 0°), which will alter the predicted angular response (Figure 1D, 1F). Overall, however, analytical and empirical angular response measurements for the PCI-borosilicate stack correspond well. Both show a nonideal angular response with cosine error magnitude >10% for the majority of angles between 0°-90° and thus do not meet the angular response design specification. Negative cosine error at all non-normal angles of incidence means that the PCI-borosilicate stack underestimates UV-C dose, though to different amounts depending on angle.

**Diffuse attenuators cause less deviation from ideal angular response**

Materials like borosilicate that exhibit specular reflection and transmittance highlight a fundamental tradeoff between extending the PCI dynamic range and minimizing cosine error. In contrast, diffuse materials are predicted to overcome this tradeoff by reducing angle-dependent reflectance (surface diffusers) and/or optical path length (volume diffusers). Available in numerous thicknesses and sizes at relatively low cost as compared to glass diffusers, PTFE is a readily available attenuator material appropriate to a wide range of environments. As a volume diffuser[37], we hypothesized that bulk scattering within PTFE would reduce path length dependence on angle of incidence. Due to unspecified surface roughness, we could not assume ideal surface diffuser behavior; thus, we modeled PTFE analytically as a volume diffuser with specular reflection and transmission at the interfaces (Figure 2D).

To assess the accuracy of the volume diffuser analytical model and characterize the extent to which PTFE alters PCI angular response, we compared both the analytical and empirical angular response of a PCI-PTFE stack to an ideal response (Figure 2E-F). For UV-C angles of incidence ≤75°, we predicted analytically and measured an integrated cosine error of 2.7% and 0.97%, respectively. Both the analytical and empirical integrated cosine errors of the PCI-PTFE stack are smaller in magnitude than observed for the PCI-borosilicate stack, as anticipated, and are lower than others have measured for 0.5 mm-thick PTFE[33,40]. We hypothesize that the lower integrated cosine error observed here could arise from differing limits of integration. Due to the limited number of angles of incidence characterized empirically, we integrate through 75°, while others[33,40] integrate through 85° (past the ISO/CIE 19476 definition[32]), incorporating contributions from an additional 10° over which cosine error is typically large. At each rotation angle measured except 90°, PCI-PTFE angular response was within error of the ideal response (Figure 2F), suggesting a near-ideal angular response. Empirical angular response was within error of model predictions at <60º; at ≥60º, the empirical PCI-PTFE stack angular response was closer to an ideal response than model predictions (Figure S6C-D). We hypothesize that the empirical angular response of the PCI-PTFE stack was closer to ideal due to some surface diffuser behavior at the interface (not incorporated in the model), and/or slight curvature or non-negligible spacing between the deformable PTFE and PCI. Diffuser-sensor spacing and diffuser curvature have been shown to substantially alter the angular response of radiometers[40–42].

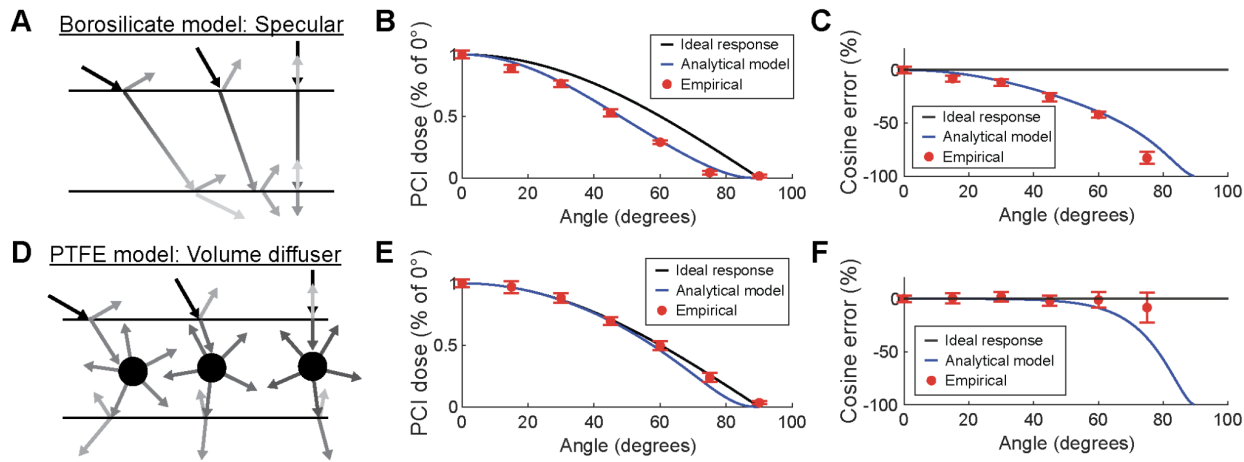

**Figure 2. Concordance of analytical and empirical angular response of PCIs stacked with specular and diffuse attenuator materials.** Analytical and empirical angular response and cosine error are compared for PCIs stacked behind (A-C) borosilicate, a model non-diffuse material, and (D-F) PTFE, a model volume diffuser. (A) Analytically, both reflection at the attenuator interfaces and path-length-dependent absorption through the material thickness contribute to the modeled angular response of non-diffuse materials. The (B) angular response and (C) cosine error of PCI-borosilicate stacks shows a non-ideal angular response at all angles of incidence. (D) The analytical model for PTFE as a volume diffuser includes specular reflection at interfaces, but assumes constant path length (and thus, absorption) through the material for all angles of incidence. The (E) angular response and (F) cosine error of PCI-PTFE stacks illustrate near-ideal response at low angles of incidence and non-ideal angular response at high angles of incidence. Error bars indicate total error, comprising both the standard deviation of replicates and the uncertainty of PCI measurements.

## Quantifying error in on-N95 UV-C dose measurements by PCI-attenuator stacks

Based on the modeled and measured angular response measurements from the point-like UV-C source, we hypothesized that a PCI-PTFE stack would measure on-N95 dose more accurately than a PCI-borosilicate stack, particularly at on-N95 locations with high angles of incidence. To test this hypothesis, we compared UV-C dose measured with PCIs and PCI-attenuator stacks to true applied dose at two locations on an N95 centered in a chamber with 5 UV-C bulbs arrayed across the top. The presence of multiple UV-C bulbs, as well as scattering and reflection[43] in this and other commercial decontamination systems, stymie determination of angle of incidence distribution at any given location. We chose two on-N95 measurement locations which we hypothesized receive substantially different angles of incidence: (1) near the apex ("low-angle"; near-normal), and (2) near the base ("high-angle"; non-normal) (Figure 3A). Based on the analytical model and the point-like UV-C source measurements (Figure 2), we hypothesized that the PCI-borosilicate stack would underestimate UV-C dose at both N95 locations, with greater underestimation at the high angle location. In contrast, PCI-PTFE angular response had cosine error magnitudes <10% at all angles of incidence measured empirically and at angles $\leq 61°$ analytically, so we hypothesized that the PCI-PTFE stack would measure on-N95 UV-C dose accurately, with some error introduced at the high-angle N95 location.

At both on-N95 locations, UV-C dose was measured from PCI color change using PCI-attenuator-specific calibration curves. (Figure 1G). To evaluate the measurement accuracy, the true dose applied at each on-N95 location was determined by multiplying a radiometer measurement obtained in each exposure by the respective predetermined ratio of irradiances at each on-N95 location and at the radiometer ($\frac{Irr_{low-angle}}{Irr_{radiometer}} = 2.27 \pm 0.06$; $\frac{Irr_{high-angle}}{Irr_{radiometer}} = 0.93 \pm 0.03$) (Note S1). Based on the ULOQ of the two PCI-attenuator stacks, on-N95 UV-C dose measurements up to ~1.200 J/cm$^2$ were characterized and compared to the true dose to evaluate the on-N95 dynamic range and angular response of PCI, PCI-borosilicate, and PCI-PTFE (Figure 3B-D). In agreement with the dynamic ranges measured on a planar surface (Figure 1G), the measured UV-C dose of the PCI-attenuator stacks is linearly proportional to true dose throughout the entire dose range tested at each on-N95 location (~0.050 to ~1.200 J/cm$^2$, Figure 3C-D, top; Table S1). Thus, both borosilicate and PTFE meet the design specification of extending on-N95 PCI dynamic range to ≥1.0 J/cm$^2$. In contrast, UV-C dose measured by the bare PCI plateaus with measurement error greatly exceeding 10% at true doses above ~0.250 J/cm$^2$ (Figure 3B), in agreement with the PCI ULOQ (Figure 1G).

To evaluate overall measurement accuracy, we calculated the percent error of UV-C dose measurements (Figure 3B-D, bottom). Doses measured with the PCI-borosilicate stack underestimated the true dose by 14.7% ± 4.0% and 40.8% ± 3.0% at the low-angle and high-angle on-N95 locations, respectively (errors are the standard deviation of 18 total dose measurements at a given location). Thus, in agreement with our hypothesis, we found that dose measured with the PCI-borosilicate stack underestimated true UV-C dose to a greater extent at the more steeply sloped, high-angle on-N95 location. Inaccuracy in measured dose also arises due to differences in the distribution of angles of incidence between the calibration curve and on-N95 measurements. As discussed, it is infeasible to generate calibration curves or correction factors specific to each on-N95 location in the chamber. In contrast, doses measured with the PCI-PTFE stack only underestimated the true dose by 3.6% ± 6.7% and 19.8% ± 5.8% at the low-angle and high-angle on-N95 locations, respectively. UV-C dose measurements by the PCI-PTFE stack were more accurate than those by the PCI-borosilicate stack, supporting our hypothesis and model predictions that PCIs stacked behind diffuse materials have an angular response nearer to an ideal response than when stacked behind a non-diffuse material. Overall, PCI-PTFE dose measurements were within error of the true dose at the low-angle on-N95 location (measured dose underestimated true dose by 3.6% ± 6.7% over 18 measurements), in agreement with our hypothesis that PCI-PTFE has near-ideal angular response at low angles of incidence. We observe greater error in PCI-PTFE-measured dose at the high-angle on-N95 location than observed at all angles measured with the point-like UV-C source (Figure 2F). The larger error at the high-angle location on-N95 may indicate an average angle of incidence >75º at that location, yielding a greater cosine error than measured with the point-like UV-C source at angles ≤75º. As discussed previously, geometrical factors such as slight variations in PTFE curvature, as well as the use of calibration curves not specific to each experimental measurement location, may have also contributed to angular response differences measured in the two systems. On-N95, the PCI-PTFE attenuator stack underestimated dose to a greater extent with increasing dose, a phenomenon not observed with the PCI-borosilicate stack (Figure 3C-D). We hypothesize the dose-dependent error may arise from an increasing difference between the true and applied calibration curve at higher doses (Figure S7), and/or temperature-induced changes in PTFE transmittance[44] not captured in the PCI-PTFE calibration curve (generated off-N95) due to differences in heat dissipation on-N95.

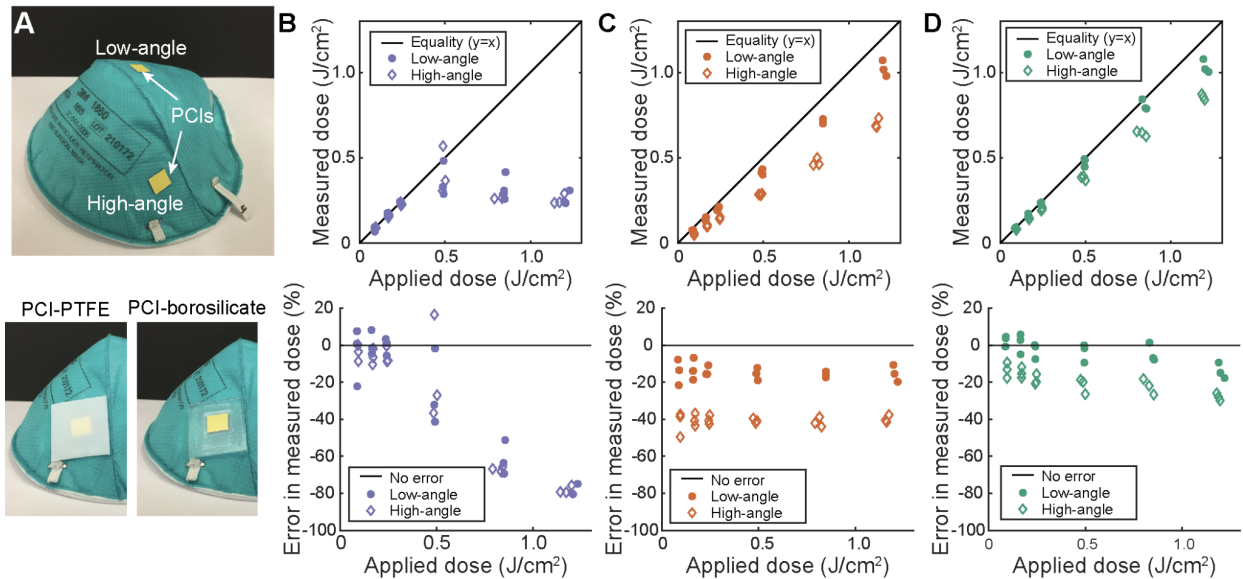

**Figure 3. On-N95 UV-C dose measurement error depends on attenuator and on-N95 location.** (A) UV-C dose was measured at two different on-N95 positions (top image): near the apex ("low-angle"), and on the steeply sloped side ("high-angle"). For PCI-attenuator stacks (PTFE or borosilicate), a PCI was placed directly underneath an attenuator (bottom image). On-N95 UV-C dose measurement accuracy of a (B) bare PCI, (C) PCI-borosilicate stack, or (D) PCI PTFE-stack was determined by comparing measured to true applied dose calculated from radiometer measurements and the predetermined ratio between the irradiance at the ratiometer and at each on-N95 location. Measured dose (top) and percent error in measured dose (bottom) were plotted against true applied dose. UV-C dose measurements underestimate true applied dose, particularly at the high-angle location.

Overall, both modeling and measurements in two different UV-C systems demonstrate that diffuse attenuators such as PTFE alter the ideal angular response of PCIs less than non-diffuse materials such as borosilicate, but that both types of planar attenuators cause deviation from ideal at high angles of incidence. Unless the material is ideally diffuse, the factors which decrease attenuator transmittance (thus increasing PCI-attenuator ULOQ) also increase the angular dependence of transmittance, yielding a fundamental tradeoff between the two design requirements of increased dynamic range and minimal cosine error. Both attenuators increased the PCI ULOQ by >4×, but the non-ideal angular response of PCI-attenuator stacks led to underestimation of measured on-N95 dose at one or both locations. The on-N95 results highlight a critical consideration for designing optical attenuators: materials that lead to measurements within error of the ideal angular response in a controlled setting may not accurately translate to user environments. Additionally, cumulative UV-C exposure also affects the transmission properties of some attenuators (e.g., solarization of glass[45]), which limits reuse. Though relatively low-cost materials such as PTFE may be feasible for single-use applications, the stability of attenuator transmittance with increasing cumulative UV-C dose must be robustly characterized prior to implementation of any attenuator material. Future study could consider introducing surface roughness and/or curvature to volume diffusers to create PCI-attenuator stacks with smaller cosine error at higher angles of incidence. Alternative strategies to extend PCI dynamic range, such as the development of new PCI formulations, are also a promising approach that may be more robust than physically attenuating UV-C incident on the PCI.

## Acknowledgments
12

We greatly appreciate the generous donation of: N95 respirators (Kristen Bole, University of California, San Francisco; UCSF), XL-1000 UV-C chamber (Betty Cheng, Spectro-UV), and the EF-140 handheld UV-C lamp (Daniel Tristan, Spectronics). We thank Prof. Aaron Streets' lab at the University of California, Berkeley (UC Berkeley) for lending the rotation platform (Thorlabs QRP02). We thank Clarence Zarobila and Dr. Cameron Miller (National Institute of Standards and Technology), Gustavo García (Spectro-UV), Henry Pinkard (UC Berkeley), and Prof. Bo Huang (UCSF) for helpful discussion.

We acknowledge the following funding sources: UC Berkeley College of Engineering Dean's COVID-19 Emergency Research Fund (PI: A.E.H), Chan Zuckerberg Biohub Investigator Program (PI: A.E.H.), NIH training grant T32GM008155 (A.G. and A.S.), National Defense Science and Engineering Graduate (NDSEG) Fellowship (A.G.), National Science Foundation Graduate Research Fellowship Program (NSF GRFP, A.S.), UC Berkeley Siebel Scholarship (A.S.), and the Natural Sciences and Engineering Research Council of Canada (NSERC) postdoctoral fellowship (S.M.G).

**References**

(1) Rutala, W. A.; Weber, D. J.; HICPAC. Guideline for Disinfection and Sterilization in Healthcare Facilities. **2019**.
(2) Memarzadeh, F.; Olmsted, R. N.; Bartley, J. M. Applications of Ultraviolet Germicidal Irradiation Disinfection in Health Care Facilities: Effective Adjunct, but Not Stand-Alone Technology. *Am J Infect Control* **2010**, *38* (5 Suppl 1), S13-24. https://doi.org/10.1016/j.ajic.2010.04.208.
(3) Tseng, C.-C.; Li, C.-S. Inactivation of Viruses on Surfaces by Ultraviolet Germicidal Irradiation. *J. Occup. Environ. Hyg.* **2007**, *4* (6), 400–405. https://doi.org/10.1080/15459620701329012.
(4) Cadnum, J. L.; Tomas, M. E.; Sankar, T.; Jencson, A.; Mathew, J. I.; Kundrapu, S.; Donskey, C. J. Effect of Variation in Test Methods on Performance of Ultraviolet-C Radiation Room Decontamination. *Infect. Control Hosp. Epidemiol.* **2016**, *37* (5), 555–560. https://doi.org/10.1017/ice.2015.349.
(5) Lowe, J. J.; Paladino, K. D.; Farke, J. D.; Boulter, K.; Cawcutt, K.; Emodi, M.; Gibbs, S.; Hankins, R.; Hinkle, L.; Micheels, T.; Schwedhelm, S.; Vasa, A.; Wadman, M.; Watson, S.; Rupp, M. E. *N95 Filtering Facepiece Respirator Ultraviolet Germicidal Irradiation (UVGI) Process for Decontamination and Reuse*; University of Nebraska Medical Center, 2020.
(6) Ozog, D. M.; Sexton, J. Z.; Narla, S.; Pretto-Kernahan, C. D.; Mirabelli, C.; Lim, H. W.; Hamzavi, I. H.; Tibbetts, R. J.; Mi, Q.-S. The Effect of Ultraviolet C Radiation Against Different N95 Respirators Inoculated with SARS-CoV-2. *Int. J. Infect. Dis.* **2020**, 224–229. https://doi.org/10.1016/j.ijid.2020.08.077.
(7) Kowalski, W. *Ultraviolet Germicidal Irradiation Handbook: UVGI for Air and Surface Disinfection*; Springer, Berlin, Heidelberg: Berlin, Heidelberg, 2009. https://doi.org/10.1007/978-3-642-01999-9.
(8) Fisher, E. M.; Shaffer, R. E. A Method to Determine the Available UV-C Dose for the Decontamination of Filtering Facepiece Respirators: UV-C Decontamination of Respirators. *J. Appl. Microbiol.* **2011**, *110* (1), 287–295. https://doi.org/10.1111/j.1365-2672.2010.04881.x.
(9) Huber, T.; Goldman, O.; Epstein, A. E.; Stella, G.; Sakmar, T. P. Principles and Practice for SARS-CoV-2 Decontamination of N95 Masks with UV-C. *Biophys. J.* **2021**, S0006349521001971. https://doi.org/10.1016/j.bpj.2021.02.039.
(10) Grist, S. M.; Geldert, A.; Gopal, A.; Su, A.; Balch, H. B.; Herr, A. E. Current Understanding of Ultraviolet-C Decontamination of N95 Filtering Facepiece Respirators. *Applied Biosafety* **2021**. https://doi.org/10.1089/apb.20.0051.

(11) Su, A.; Grist, S. M.; Geldert, A.; Gopal, A.; Herr, A. E. Quantitative UV-C Dose Validation with Photochromic Indicators for Informed N95 Emergency Decontamination. *PLoS ONE* **2021**, *16* (1), e0243554. https://doi.org/10.1371/journal.pone.0243554.

(12) Lawal, O.; Dussert, B.; Howarth, C.; Platzer, K.; Sasges, M.; Muller, J.; Whitby, E.; Bolton, J.; Photosciences, B.; Santelli, M. Method for the Measurement of the Output of Monochromatic (254 Nm) Low-Pressure UV Lamps. *IUVA News* **2017**, *19* (1).

(13) Reifsnyder, W. E. Radiation Geometry in the Measurement and Interpretation of Radiation Balance. *Agricultural Meteorology* **1967**, *4* (4), 255–265. https://doi.org/10.1016/0002-1571(67)90026-X.

(14) McCluney, W. R. *Introduction to Radiometry and Photometry, Second Edition*; Artech House, 2014.

(15) Zorzano, M.-P.; Martín-Soler, J.; Gómez-Elvira, J. UV Photodiodes Response to Non-Normal, Non-Colimated and Diffusive Sources of Irradiance. In *Photodiodes - Communications, Bio-Sensings, Measurements and High-Energy Physics*; IntechOpen, 2011.

(16) Larason, T.; Ohno, Y. Calibration and Characterization of UV Sensors for Water Disinfection. *Metrologia* **2006**, *43* (2), S151–S156. https://doi.org/10.1088/0026-1394/43/2/S30.

(17) Geldert, A.; Su, A.; Roberts, A. W.; Golovkine, G.; Grist, S. M.; Stanley, S. A.; Herr, A. E. Nonuniform UV-C Dose across N95 Facepieces Can Cause 2.9-Log Variation in SARS-CoV-2 Inactivation. *medRxiv* **2021**. https://doi.org/10.1101/2021.03.05.21253022.

(18) Mills, A.; McDiarmid, K.; McFarlane, M.; Grosshans, P. Flagging up Sunburn: A Printable, Multicomponent, UV-Indicator That Warns of the Approach of Erythema. *Chem. Commun.* **2009**, No. 11, 1345–1346. https://doi.org/10.1039/b900569b.

(19) Smith, A. T.; Ding, H.; Gorski, A.; Zhang, M.; Gitman, P. A.; Park, C.; Hao, Z.; Jiang, Y.; Williams, B. L.; Zeng, S.; Kokkula, A.; Yu, Q.; Ding, G.; Zeng, H.; Sun, L. Multi-Color Reversible Photochromisms via Tunable Light-Dependent Responses. *Matter* **2020**, *2* (3), 680–696. https://doi.org/10.1016/j.matt.2019.12.006.

(20) Khiabani, P. S.; Soeriyadi, A. H.; Reece, P. J.; Gooding, J. J. Paper-Based Sensor for Monitoring Sun Exposure. *ACS Sens.* **2016**, *1* (6), 775–780. https://doi.org/10.1021/acssensors.6b00244.

(21) Parisi, A. V.; Kimlin, M. G. Personal Solar UV Exposure Measurements Employing Modified Polysulphone with an Extended Dynamic Range. *Photochem. Photobiol.* **2007**, *79* (5), 411–415. https://doi.org/10.1111/j.1751-1097.2004.tb00028.x.

(22) Furler, R. A. Angular Dependence of Optical Properties of Homogeneous Glasses. *ASHRAE Transactions* **1991**, *97*, 1–9.

(23) Seckmeyer, G.; Bernhard, G. Cosine Error Correction of Spectral UV Irradiance. *SPIE Proc.* **1993**, *2049*. https://doi.org/10.1117/12.163505.

(24) Severin, B. F.; Roessler, P. F. Resolving UV Photometer Outputs with Modeled Intensity Profiles. *Water Research* **1998**, *32* (5), 1718–1724. https://doi.org/10.1016/S0043-1354(97)00363-1.

(25) SCHOTT Technical Glass Solutions GmbH. BOROFLOAT® 33 – Optical Properties https://www.schott.com/d/borofloat/bde16ad3-70e5-48a0-b8ac-9146fcd34511/1.0/borofloat33_opt_en_web.pdf (accessed Jul 7, 2020).

(26) Lake Photonics. Spectralex Optical Diffuser Films https://www.lake-photonics.com/wp-content/uploads/Spectralex-Optical-Diffuser-Films.pdf (accessed May 6, 2021).

(27) Luo, M. R.; Cui, G.; Rigg, B. The Development of the CIE 2000 Colour-Difference Formula: CIEDE2000. *Color Res. Appl.* **2001**, *26* (5), 340–350. https://doi.org/10.1002/col.1049.

(28) Quintern, L. E.; Horneck, G.; Eschweiler, U.; Bücker, H. A Biofilm Used as Ultraviolet-Dosimeter. *Photochemistry and Photobiology* **1992**, *55* (3), 389–395. https://doi.org/10.1111/j.1751-1097.1992.tb04252.x.

(29) Lore, M. B.; Heimbuch, B. K.; Brown, T. L.; Wander, J. D.; Hinrichs, S. H. Effectiveness of Three Decontamination Treatments against Influenza Virus Applied to Filtering Facepiece Respirators. *Ann. Occup. Hyg.* **2011**, *56* (1), 92–101. https://doi.org/10.1093/annhyg/mer054.

(30) Mills, D.; Harnish, D. A.; Lawrence, C.; Sandoval-Powers, M.; Heimbuch, B. K. Ultraviolet Germicidal Irradiation of Influenza-Contaminated N95 Filtering Facepiece Respirators. *Am. J. Infect. Control* **2018**, *46* (7), e49–e55. https://doi.org/10.1016/j.ajic.2018.02.018.

(31) Heimbuch, B.; Harnish, D. *Research to Mitigate a Shortage of Respiratory Protection Devices During Public Health Emergencies*; Report to the FDA HHSF223201400158C; Applied Research Associate, Inc, 2019.

(32) *Characterization of the Performance of Illuminance Meters and Luminance Meters*; ISO/CIE 19476:2014; International Organization for Standardization/International Commission on Illumination, 2014.

(33) Bernhard, G.; Seckmeyer, G. New Entrance Optics for Solar Spectral UV Measurements. *Photochem. Photobiol.* **1997**, *65* (6), 923–930. https://doi.org/10.1111/j.1751-1097.1997.tb07949.x.

(34) Allen, D. W.; Early, E. A.; Tsai, B. K.; Cooksey, C. C. *NIST Measurement Services: Regular Spectral Transmittance*; NIST SP 250-69; National Institute of Standards and Technology: Gaithersburg, MD, 2011. https://doi.org/10.6028/NIST.SP.250-69.

(35) Miller, D. C.; Kempe, M. D.; Kennedy, C. E.; Kurtz, S. R. Analysis of Transmitted Optical Spectrum Enabling Accelerated Testing of Multijunction Concentrating Photovoltaic Designs. *OE* **2011**, *50* (1), 013003. https://doi.org/10.1117/1.3530092.

(36) Li, Q.; Lee, B. J.; Zhang, Z. M.; Allen, D. W. Light Scattering of Semitransparent Sintered Polytetrafluoroethylene Films. *J. Biomed. Opt.* **2008**, *13* (5), 054064. https://doi.org/10.1117/1.2992485.

(37) Lemaillet, P.; Patrick, H. J.; Germer, T. A.; Hanssen, L.; Johnson, B. C.; Georgiev, G. T. Goniometric and Hemispherical Reflectance and Transmittance Measurements of Fused Silica Diffusers; Hanssen, L. M., Ed.; San Diego, California, United States, 2016; p 996109. https://doi.org/10.1117/12.2237975.

(38) Pye, S. D.; Martin, C. J. A Study of the Directional Response of Ultraviolet Radiometers: I. Practical Evaluation and Implications for Ultraviolet Measurement Standards. *Phys. Med. Biol.* **2000**, *45* (9), 2701–2712. https://doi.org/10.1088/0031-9155/45/9/319.

(39) Sholtes, K. A.; Lowe, K.; Walters, G. W.; Sobsey, M. D.; Linden, K. G.; Casanova, L. M. Comparison of Ultraviolet Light-Emitting Diodes and Low-Pressure Mercury-Arc Lamps for Disinfection of Water. *Environmental Technology* **2016**, *37* (17), 2183–2188. https://doi.org/10.1080/09593330.2016.1144798.

(40) Pulli, T.; Kärhä, P.; Mes, J.; Schreder, J.; Jaanson, P.; Manoocheri, F. Improved Diffusers for Solar UV Spectroradiometers. *AIP Conference Proceedings* **2013**, *1531* (1), 813–816. https://doi.org/10.1063/1.4804894.

(41) Martínez, M. A.; Andújar, J. M.; Enrique, J. M. A New and Inexpensive Pyranometer for the Visible Spectral Range. *Sensors (Basel)* **2009**, *9* (6), 4615–4634. https://doi.org/10.3390/s90604615.

(42) Cahuantzi, R.; Buckley, A. Geometric Optimisation of an Accurate Cosine Correcting Optic Fibre Coupler for Solar Spectral Measurement. *Rev. Sci. Instrum.* **2017**, *88* (9), 095003. https://doi.org/10.1063/1.5003040.

(43) Gostein, M.; Stueve, B. Accurate Measurement of UV Irradiance in Module-Scale UV Exposure Chambers, Including Spectral Angular Response of Sensor. In *2016 IEEE 43rd Photovoltaic Specialists Conference (PVSC)*; 2016; pp 0863–0866. https://doi.org/10.1109/PVSC.2016.7749731.

(44) Ylianttila, L.; Schreder, J. Temperature Effects of PTFE Diffusers. *Optical Materials* **2005**, *27* (12), 1811–1814. https://doi.org/10.1016/j.optmat.2004.11.008.

(45) Gatto, A.; Escoubas, L.; Roche, P.; Commandré, M. Simulation of the Degradation of Optical Glass Substrates Caused by UV Irradiation While Coating. *Opt. Commun.* **1998**, *148* (4), 347–354. https://doi.org/10.1016/S0030-4018(97)00651-2.

# Supporting Information

## Optical attenuators extend dynamic range but alter angular response of planar ultraviolet-C dosimeters

Alison Su[1#], Alisha Geldert[1#], Samantha M. Grist[2], & Amy E. Herr[1,2*]

[#]These authors contributed equally.

*Corresponding author: aeh@berkeley.edu

[1]University of California, Berkeley-University of California, San Francisco Graduate Program in Bioengineering, [2]Department of Bioengineering, Berkeley, California 94720, United States

## Table of Contents

**Figure S1.** Measurement setups to characterize borosilicate transmittance and PCI-attenuator stack angular response

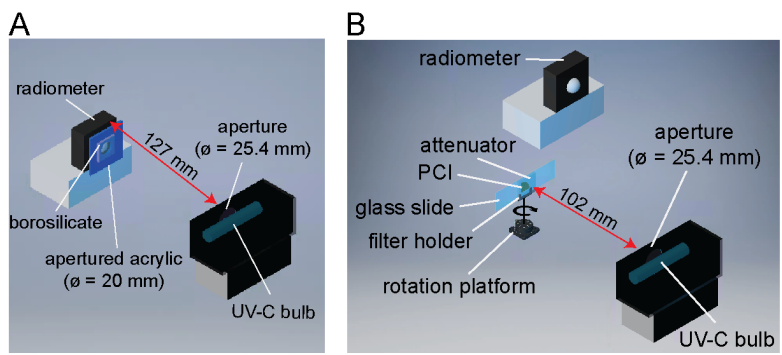

**Figure S1. Schematics of measurement setups to characterize borosilicate transmittance and PCI-attenuator stack angular response.** (A) Borosilicate transmittance ($T_{total}$) at near-normal angles of incidence is measured by comparing irradiance measurements with and without borosilicate in the optical path. To ensure borosilicate is placed normal to the optical path and radiometer, borosilicate is mounted on a custom-made acrylic (McMaster-Carr 85635K421) platform with a 20 mm-diameter aperture centered over the radiometer sensor (borosilicate is placed ~9 mm in front of the top of the radiometer diffuser dome). The acrylic blocks all UV-C, so UV-C is incident only through the 20 mm aperture. For homogeneous materials, the attenuation coefficient (α) can be calculated from the measured $T_{total}$ and modeled $T_{int}$ at 0°, and the attenuator thickness (d):

$$\alpha = \frac{-ln(T_{total}(0°)/(T_{int1}(0°)T_{int2}(0°)))}{d} = \frac{-ln(T_{mat}(0°))}{d}$$

(B) Angular response of PCI-attenuator stacks is measured by exposing the PCI-attenuator stack to UV-C from a point-like source at different angles of incidence. PCI-attenuator stacks were affixed to a glass microscope slide (VWR 48300-026) with double-sided tape (3M MMM137). The glass slide was held in a filter holder (Thorlabs FH2) on an optical post attached to a rotation platform (Thorlabs QRP02). Arrow around optical post denotes axis of rotation. We assumed the borosilicate and PCI-attenuator stacks received negligible reflected and scattered UV-C, as no enclosure was used and wall paint is minimally UV-C-reflective.[1]

**Figure S2.** UV-C chamber floor map for calibration curve measurements

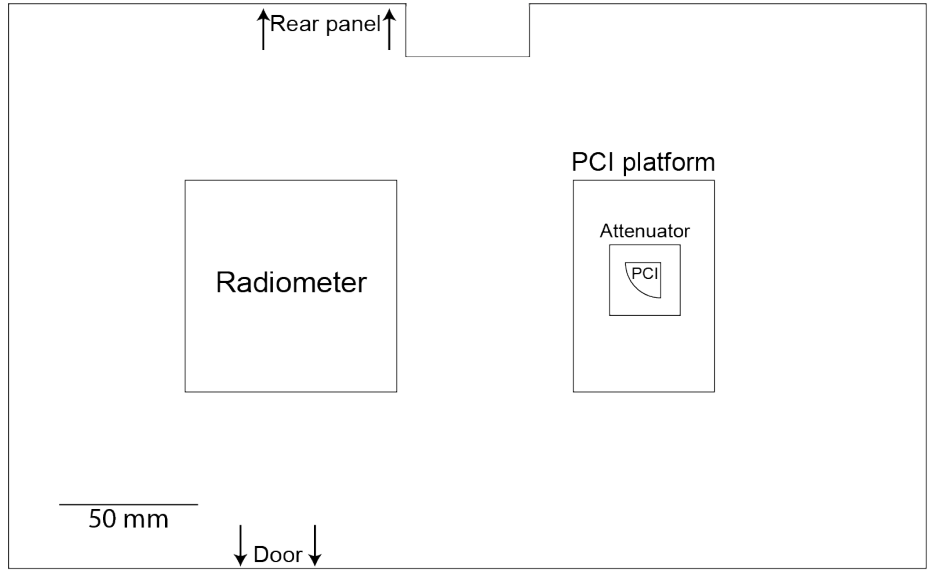

**Figure S2. UV-C chamber floor map for calibration curve measurements.** The PCI was placed on a custom acrylic platform to raise the PCI to the same height as the base of the Teflon dome on the radiometer. For PCI-attenuator stack calibration curves, the attenuator was placed directly on top of the PCI on the platform. Irradiances at the radiometer and PCI locations were verified to be equivalent. Rectangular cut-out near the rear panel allows the floor map to fit around a raised component built into the UV-C chamber.

**Figure S3.** UV-C chamber floor map for on-N95 measurements

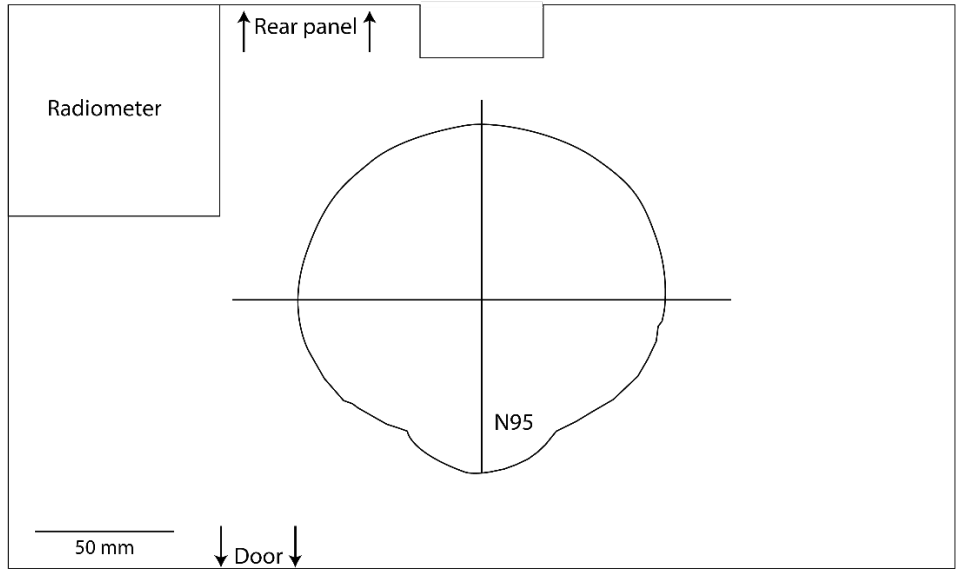

**Figure S3. UV-C chamber floor map for on-N95 measurements.** Dose measurements from the radiometer were used to determine the true dose applied to the N95 surface based on the predetermined irradiance ratio between the radiometer and each on-N95 location.

**Figure S4.** Attenuators extend PCI dynamic range

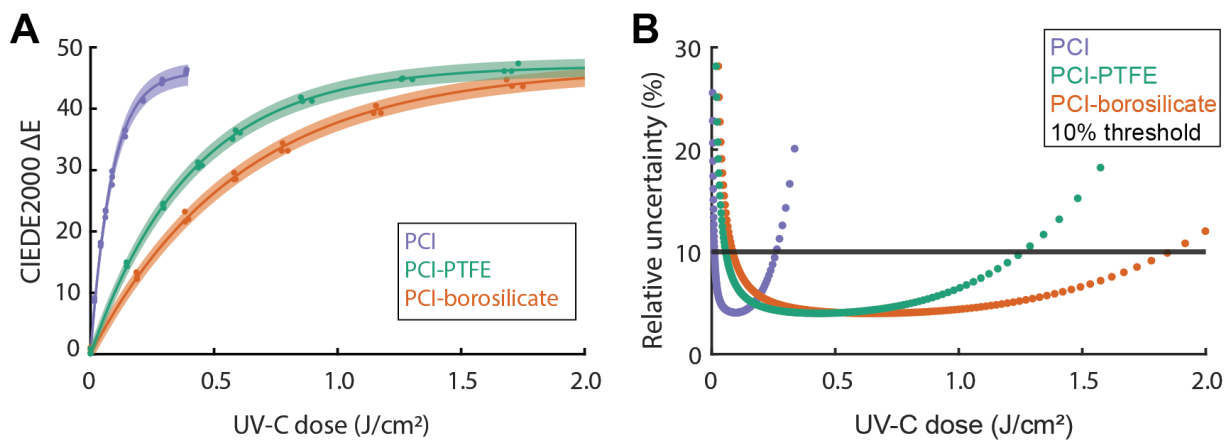

**Figure S4. Attenuators extend PCI dynamic range.** (A) Calibration curves relating UV-C dose to PCI color change (CIEDE2000 ΔE). UV-C was applied to either bare PCIs or PCIs stacked directly underneath an attenuator (0.51 mm-thick PTFE or 1.1 mm-thick borosilicate). Based on first-order reaction kinetics, a fit function ($\Delta E = a\{1 - e^{-\frac{dose}{b}}\}$) is defined for each calibration curve, as described previously[2]. Shaded regions indicate the 95% prediction interval on prediction of PCI color change from measured UV-C dose. For bare PCIs, $R^2$ = 0.9976, a = 46.0 (45.3, 46.7), b = 87.4 (83.6, 91.2). For PCI-PTFE, $R^2$ = 0.9982, a = 47.0 (46.5, 47.5), b = 407.3 (393.2, 421.4). For PCI-borosilicate, $R^2$ = 0.9982, a = 46.7 (46,2, 47.3), b = 605.9 (584.3, 627.5). For each calibration curve, N = 3 replicates were measured at each target dose. (B) The dynamic range (LLOQ to ULOQ) is defined as the dose range over which relative uncertainty in dose measurement is <10%. Relative uncertainty is defined as half the width of the 95% confidence interval on UV-C dose measurements, divided by measured dose. UV-C dose measurements have <10% relative uncertainty from 0.011 – 0.261 $J/cm^2$ (bare PCI), 0.057 – 1.259 $J/cm^2$ (PCI-PTFE), and 0.085 – 1.853 $J/cm^2$ (PCI-borosilicate).

**Figure S5.** Borosilicate transmittance measurement involves a maximum angle of incidence of 10.1° from apertured UV-C source

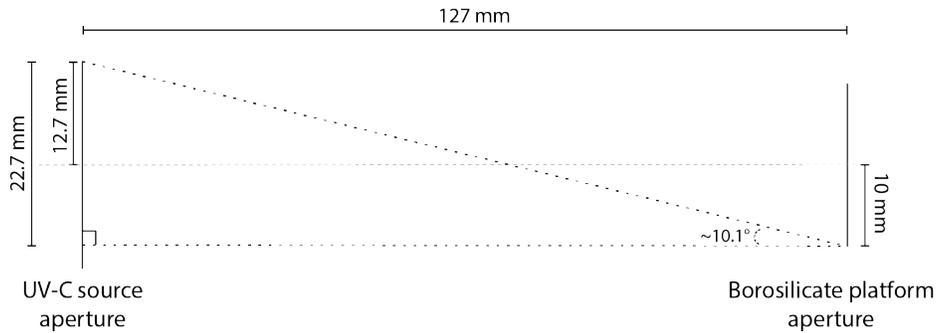

**Figure S5. Borosilicate transmittance measurement involves a maximum angle of incidence of 10.1° from apertured UV-C source.** To measure the total transmittance through borosilicate at near-normal incidence, the maximum angle incident on the borosilicate should be minimized. Borosilicate is placed on an apertured platform in front of the radiometer. In our setup, the maximum angle incident on the borosilicate from the apertured UV-C source is ~10.1°.

**Figure S6.** Analytical and empirical angular responses of PCI-attenuator pairs are concordant

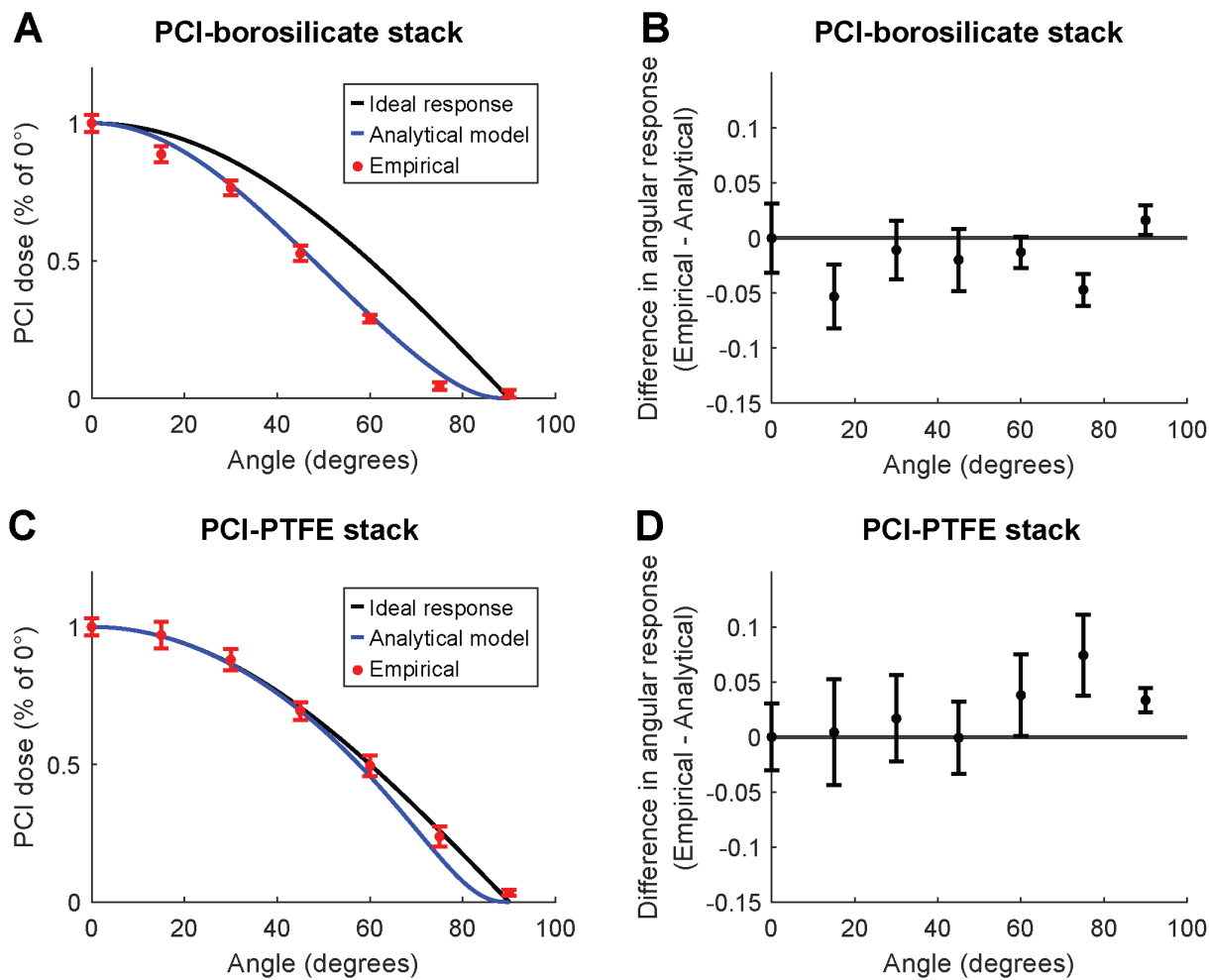

**Figure S6. Analytical and empirical angular responses of PCI-attenuator pairs are concordant.** (A) Analytical and empirical angular response of the PCI-borosilicate stack, along with ideal angular response (cos(θ)). Error bars on empirical measurements indicate total propagated error (the root-sum-square combination of both PCI quantification uncertainty and standard deviation of 3 replicate measurements, as described previously[2]). (B) The difference between empirical and analytical angular response of the PCI-borosilicate stack at each angle of incidence measured. (C) Analytical and empirical angular response of the PCI-PTFE stack, along with ideal angular response. (D) The difference between empirical and analytical angular response of the PCI-PTFE stack at each angle of incidence measured.

**Figure S7.** Use of incorrect calibration curve can yield dose-dependent measurement error

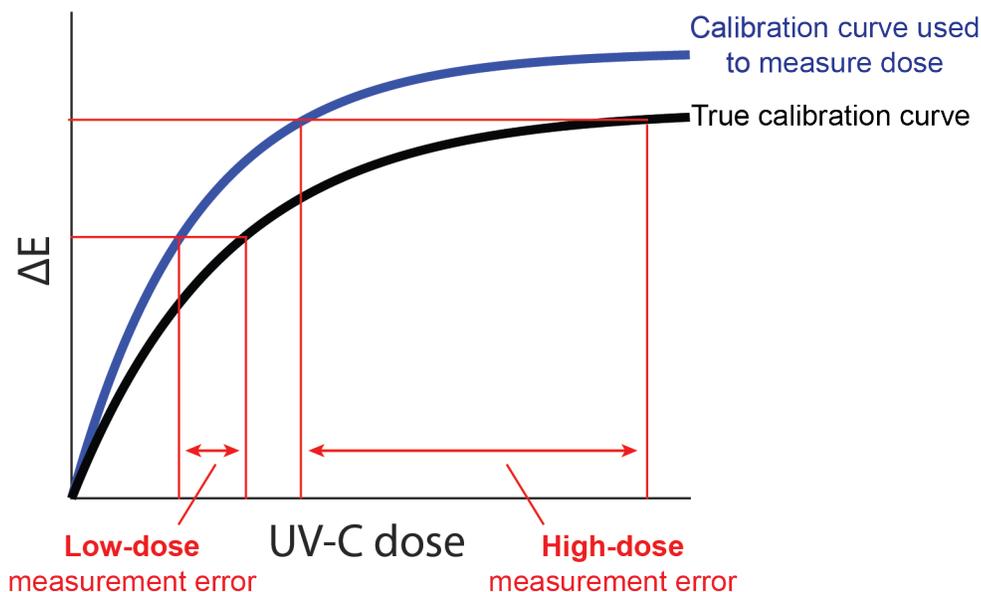

**Figure S7. Use of incorrect calibration curve can yield dose-dependent measurement error.** If the PCI-attenuator stack has a non-ideal angular response, and/or if PCI reaction kinetics are dependent on environmental conditions such as temperature or humidity, differences in UV-C angles of incidence and environmental factors may yield different calibration curve shapes. Because it is infeasible to generate calibration curves for every location and environmental condition within the UV-C chamber to exactly match the conditions of a given measurement, the calibration curve used to determine dose from a measured ΔE may not represent the true calibration curve for the exact chamber location and environmental conditions present at the time the PCI was exposed. Use of an incorrect calibration curve may lead to dose-dependent measurement error.

**Table S1.** Significance of linear correlation between true and measured dose for each attenuator and on-N95 location tested

| Attenuator & on-N95 location | $r^2$ | p |
|---|---|---|
| No attenuator, low-angle | 0.282 | 0.024 |
| No attenuator, high-angle | 0.182 | 0.078 |
| Borosilicate, low-angle | 0.997 | 4.76e-21 |
| Borosilicate, high-angle | 0.997 | 8.21e-22 |
| PTFE, low-angle | 0.990 | 2.64e-17 |
| PTFE, high-angle | 0.995 | 1.68e-19 |

**Note S2.** Method of determining true on-N95 dose.

To calculate the true UV-C dose applied at the N95 surface when either the dose exceeded the PCI upper limit of quantification or an attenuator was used, we first determined the ratio of irradiance at each on-N95 location to the irradiance at the radiometer (n = 3 replicates). The radiometer recorded irradiance during UV-C exposure of an N95 with bare PCIs affixed at the low- and high-angle locations. Exposure time was chosen so that the dose applied to both PCIs was within the PCI dynamic range, because unmodified PCIs have an ideal angular response[3]. We assumed the irradiance ratio between locations is independent of dose, as exposure times (≤ 6 min) were substantially shorter than the timescales over which spatial variation in bulb output has been observed.[4] PCI dose was quantified as described above, and dose at the radiometer was quantified by integrating the recorded irradiance. Radiometer measurements in subsequent experiments were scaled by the ratio to calculate the true applied on-N95 dose for measurements made by PCI-attenuator stacks and PCIs exposed to UV-C doses that exceeded the bare PCI dynamic range.

# References

(1) Rutala, W. A.; Gergen, M. F.; Tande, B. M.; Weber, D. J. Rapid Hospital Room Decontamination Using Ultraviolet (UV) Light with a Nanostructured UV-Reflective Wall Coating. *Infect. Control Hosp. Epidemiol.* **2013**, *34* (5), 527–529. https://doi.org/10.1086/670211.

(2) Su, A.; Grist, S. M.; Geldert, A.; Gopal, A.; Herr, A. E. Quantitative UV-C Dose Validation with Photochromic Indicators for Informed N95 Emergency Decontamination. *PLoS ONE* **2021**, *16* (1), e0243554. https://doi.org/10.1371/journal.pone.0243554.

(3) Geldert, A.; Su, A.; Roberts, A. W.; Golovkine, G.; Grist, S. M.; Stanley, S. A.; Herr, A. E. Nonuniform UV-C Dose across N95 Facepieces Can Cause 2.9-Log Variation in SARS-CoV-2 Inactivation. *medRxiv* **2021**, 2021.03.05.21253022. https://doi.org/10.1101/2021.03.05.21253022.

(4) Schmalwieser, A. W.; Wright, H.; Cabaj, A.; Heath, M.; Mackay, E.; Schauberger, G. Aging of Low-Pressure Amalgam Lamps and UV Dose Delivery. *J. Environ. Eng. Sci.* **2014**, *9* (2), 113–124. https://doi.org/10.1680/jees.13.00009.